\documentclass[aps,preprint,groupedaddress,showpacs,fleqn]{revtex4}
\usepackage[dvips]{graphicx}
\begin{document}
\title{Lepton flavor violation 
in the supersymmetric standard model with vectorlike leptons}
\author{Ryuichiro Kitano}
\affiliation{Theory Group, KEK, 
Oho 1-1, Tsukuba, Ibaraki 305-0801, Japan}

\affiliation{Department of Particle and Nuclear Physics,
The Graduate University for Advanced Studies,
Oho 1-1, Tsukuba, Ibaraki 305-0801, Japan}

\author{Katsuji Yamamoto}
\affiliation{Department of Nuclear Engineering, Kyoto University,
Kyoto 606-8501, Japan}
\begin{abstract}
Lepton flavor violating processes are obtained
from the mixing between ordinary leptons and
vectorlike ${\rm SU(2)_{\rm L}}$ doublet leptons
which may originate in ${\rm E_6}$.
The effects of this lepton mixing are, however, suppressed naturally
by the hierarchy of the charged lepton masses.
In the supersymmetric model, significant effects of lepton flavor violation
may appear rather through slepton mixing,
which is in the present case generated by radiative corrections
with ordinary-exotic lepton couplings.
We are especially interested in the $\mu \rightarrow e \gamma$ decay.
In the model without the bare mass term of vectorlike leptons,
the supersymmetric contributions are rather suppressed
due to the approximate $ {\rm U(1)}_e \times {\rm U(1)}_\mu $.
It is, however, remarkable that they are substantially enhanced
by $ \tan^6 \beta $.
Then, $ {\rm Br} ( \mu \rightarrow e \gamma ) $ might be comparable
to the experimental bound for large $ \tan \beta $.
In the model with the bare mass term,
much larger contributions are obtained through slepton mixing.
These investigations show that the supersymmetric effects
on lepton flavor violation due to the vectorlike leptons
can be observed in the near future experiments.
\end{abstract}
\pacs{14.60.Hi, 11.30.Fs, 12.60.Jv, 13.35.Bv}
\maketitle

\section{Introduction}
The ${\rm E_6}$ supersymmetric (SUSY) grand unified theory (GUT) and
the ${\rm E_8} \times {\rm E_8^\prime}$ superstring theory
are good candidates for unified theories
of elementary particle physics.
While such theories manifest themselves at the ultrahigh energy scale
($\gtrsim 10^{16}$ GeV),
it is natural to expect that they have some effects on physics
at the electroweak scale ($\sim 10^2$ GeV)
\cite{e6}.
It is, however, difficult to determine precisely what effective theory
actually appears at the electroweak scale.
This may be understood by considering the facts
that the ${\rm E_6}$ SUSY GUT has 
many possible symmetry breaking patterns
and that the superstring theory has
many possible compactification patterns.
Hence, in the present experimental circumstances,
it will not be so useful to perform detailed investigations
on specific models.
We are rather interested in the general aspects
of the low-energy effects of the unified theories.
Particularly, in this article we study the effects on lepton physics
at the electroweak scale.
In the standard model, the lepton number and flavor are
conserved automatically.  Hence, if the processes violating
these quantum numbers are observed,
including the neutrino masses and oscillations,
$\mu \to e \gamma$, $\mu \to 3e$, and so on,
they indeed provide important information of new physics
above the electroweak scale.

In the supersymmetric models 
with supergravity scenario \cite{Chamseddine:1982jx},
which would be derived from the unified theories,
the renormalization effects on the soft supersymmetry breaking terms
may induce sizable lepton flavor violation.
These lepton flavor violating renormalization effects
are actually obtained in the SUSY GUT
\cite{Barbieri:1994pv,Barbieri:1995tw,hisano2}
and the minimal supersymmetric standard model (MSSM)
with heavy right-handed neutrinos
\cite{Borzumati:1986qx,hisano,Hisano:1998fj}.
Here, as another intriguing possibility for lepton flavor violation,
we investigate the supersymmetric standard model
incorporating vectorlike ${\rm SU(2)_{\rm L}}$ doublet leptons
\cite{ll,Dubovsky:1997rq}.
Especially, the significant slepton mixing may be generated
by radiative corrections with the couplings
between the ordinary leptons and these exotic leptons.
It is remarkable that this new physics with exotic leptons
can be discovered just around the electroweak scale
in contrast to the cases of GUT and right-handed neutrinos.

The present model is motivated by the ${\rm E_6}$ type unification.
The ${\rm E_6}$ {\bf 27} representation contains
one generation of the ordinary quarks and leptons in the standard model,
the vectorlike ${\rm SU(2)_{\rm L}}$ singlet quarks,
the vectorlike ${\rm SU(2)_{\rm L}}$ doublet leptons,
the right-handed neutrino, and the ${\rm SU(2)_{\rm L}}$ singlet Higgs field.
The ordinary left-handed neutrinos may acquire Majorana masses
through the seesaw mechanism with the right-handed neutrinos \cite{seesaw}.
These ordinary neutrino masses can be very small as required
phenomenologically, if the right-handed neutrinos are superheavy
or if they have extremely small Yukawa couplings
with the left-handed lepton doublets.
Hence, given the small ordinary neutrino masses,
we do not consider further the effects
of the right-handed neutrinos in the electroweak physics.
In this investigation, we concentrate
on the lepton flavor violating effects
of the vectorlike lepton doublets $( L , L^c )$
and their scalar superpartners.
Specifically, these supersymmetric effects of the exotic leptons
may provide the significant Br($\mu \to e \gamma$),
while the lepton flavor changing neutral currents (FCNC's)
at the tree level are suppressed naturally
by the hierarchy of the charged lepton masses.

The down type quark singlets
$ ( D^c , D ) \sim {\bar{\bf 3}}_C + {\bf 3}_C $
are also present in the ${\rm E_6}$ unification,
which may form the $ {\bar{\bf 5}} + {\bf 5} $ of SU(5) subgroup
with $ ( L , L^c ) \sim {\bf 2}_{\rm L} + {\bf 2}_{\rm L} $.
The survival of these exotic particles to the electroweak scale
depends on the symmetry breakings.
The $ {\bf 2}_{\rm L} $-$ {\bf 3}_C $ mass splitting in the {\bf 5}
can be realized with suitable Higgs multiplets,
e.g. the {\bf 351} of ${\rm E_6}$ including the {\bf 24} of SU(5),
as usually considered to obtain the light ${\rm SU(2)_L}$ doublet Higgs field.
Hence, it is possible that $ ( L , L^c ) $ survive
while $ ( D^c , D ) $ become superheavy in the {\bf 27}.
It may also be considered in the superstring theory
that $ ( L , L^c ) $ become light while $ ( D^c , D ) $ do not
in the $ {\overline{\bf 27}} + {\bf 27} $.
On the other hand, $ ( D^c , D ) $ and $ ( L , L^c ) $
are expected to survive together for the gauge coupling unification.
Even in this case, $ ( D^c , D ) $ and $ ( L , L^c ) $
may originate in different {\bf 27}'s
and/or ($ {\overline{\bf 27}} + {\bf 27} $)'s.
Then, it should be noted in these cases,
whether $ ( D^c , D ) $ survive or not,
that the simple ${\rm E_6}$ relations no longer hold
for the Yukawa couplings after the ${\rm E_6}$ breaking.
In particular, the flavor violating couplings with $ ( L , L^c ) $
may be described independently of those with $ ( D^c , D ) $.
Given this general situation, we concentrate
on the lepton flavor violation with $ ( L , L^c ) $ in the present article.
A detailed examination of the quark flavor violation with $ ( D^c , D ) $,
particularly the supersymmetric effects,
is reserved for a separate publication.
(Such supersymmetric effects are investigated in Ref.\cite{Dubovsky:1997rq}
in a context different from the present one.)
Of course, if $ ( D^c , D ) $ and $ ( L , L^c ) $
originate in the same {\bf 27},
the flavor violating effects of these exotic leptons and quarks
will be provided from some common ${\rm E_6}$ coupling.
In this specific case, the calculations made in the present analysis
for the lepton flavor violation such as $ \mu \rightarrow e \gamma $
can be extended readily for the quark flavor violation
such as $ K^0 $-$ {\bar K}^0 $ mixing.
Simple order estimates, as will be quoted occasionally in the text, indicate
that the constraints from the quark flavor violating effects are
in fact less stringent than those from the lepton flavor violating effects.

This paper is organized as follows.
In Section II, we describe the lepton mixing in the presence
of vectorlike lepton doublets.
Then, the contributions to Br($\mu \to e \gamma$)
due to the ordinary-exotic lepton mixing are shown to be negligibly small.
In Section III, we investigate the slepton mixing generated
by radiative corrections with the ordinary-exotic lepton couplings.
Significant contributions to Br($\mu \to e \gamma$)
can be obtained from this slepton mixing.
Section IV is devoted to the summary.

\section{The lepton mixing}

In this section, we describe the lepton mixing
including the vectorlike lepton doublets,
which contributes to the lepton flavor changing processes
such as $\mu \to e \gamma$.
We show that the modification of the gauge interactions
provided by this lepton mixing is fairly suppressed
even if the relevant flavor violating Yukawa couplings
take significant values.
This feature is indeed attributed to the mass hierarchy
between ordinary leptons and exotic leptons.

The ordinary leptons are represented by
$l_i \equiv (\nu_i,e_i)$ and $e_i^c$ with the subscript
$i=1,2,3$ for the generations.
The vectorlike leptons and singlet Higgs field are also listed as follows
with their quantum numbers of
the $ {\rm SU(2)_{\rm L}} \times {\rm U(1)}_Y $:
\begin{eqnarray}
L \equiv l_4 = \left( \begin{array}{c} N \\ E \end{array} \right) :
({\bf 2},-1/2) \ , \
L^c = \left( \begin{array}{c} E^c \\ N^c \end{array} \right) :
({\bf 2},1/2) \ , \
S : ({\bf 1},0) \ .
\end{eqnarray}
We here describe the case with one generation of $L$ and $L^c$
for simplicity.  It is straightforward to extend the analysis
for several generations of vectorlike leptons,
and similar results are obtained for the lepton flavor violating processes.
We examine two typical models in the following,
for which distinct results will be obtained
concerning the supersymmetric contributions
to the lepton number violating processes
such as $ \mu \rightarrow e \gamma $.

In the first model, the superpotential relevant
for the leptonic superfields is given by
\begin{eqnarray}
{\cal W} = \lambda_1^{aj} l_a H_1 e_j^c + \lambda_2^{a4} l_a L^c S
+ \lambda_3 H_1 H_2 S + \frac{\lambda_4}{3} S^3 \ ,
\label{2}
\end{eqnarray}
where $ a = i,4 $ and $ i,j = 1,2,3 $.
The Higgs fields include the pair of ${\rm SU(2)_{\rm L}}$ doublets
$H_1$ and $H_2$ and the singlet $S$.
The vacuum expectation value (VEV) of $S$ gives the mass terms
involving the exotic lepton doublets $L$ and $L^c$
as well as the $\mu$ term for the Higgs doublets $H_1$ and $H_2$.
The $R$-parity invariance is assumed for the lepton number conservation.
The ${\rm Z_3}$ symmetry is also assumed to exclude
the mass term of $S$ which is independent of the electroweak scale.
(The cosmological domain wall problem associated with the spontaneous
breaking of ${\rm Z_3}$ may be evaded by introducing
small explicit breaking in the higher order terms.)
The coupling matrices in eq.(\ref{2}) may be simplified
without loss of generality by redefinition of the relevant fields
at some unification scale such as the gravitational scale $ M_{\rm G} $.
The elements $ \lambda_1^{4j} $ in the $ 4 \times 3 $ coupling matrix
$ \lambda_1 $ can be eliminated by an appropriate unitary transformation
among the left-handed doublets $ l_i $ and $ l_4 \equiv L $.
Then, the submatrix $ \lambda_1^{ij} $ can be diagonalized
by a redefinition of the ordinary leptons.
Accordingly, we may take
\begin{eqnarray}
\lambda_1 ( M_{\rm G} ) = \left( \begin{array}{c}
\delta_{ij} \lambda_1^{ii} \\ {\bf 0} \end{array} \right) \ .
\label{3}
\end{eqnarray}
Furthermore, the components of $\lambda_1 - \lambda_4$
can all be made real and non-negative.

In the second model, the above form of $ \lambda_1 $ is realized
in terms of another ${\rm Z_3}$, under which
the nontrivial transformations of the relevant fields are given by
\begin{eqnarray}
L \rightarrow \omega L , \ L^c \rightarrow \omega^* L^c , \
S \rightarrow \omega S , \ H_2 \rightarrow \omega^* H_2
\label{4}
\end{eqnarray}
with $ \omega^3 = 1 $.
Then, we have a bare mass term for $ L $ and $ L^c $
by making the following replacement in the superpotential (\ref{2}):
\begin{eqnarray}
\lambda_2^{44} L L^c S \ {\mbox{\bf [model (1)]}} \
\rightarrow \ M_L L L^c \ {\mbox{\bf [model (2)]}} \ .
\end{eqnarray}

In the lepton basis where the $ \lambda_1 ( M_{\rm G} ) $
is given by eq.(\ref{3}), the ordinary-exotic lepton couplings
$ \lambda_2^{i4} $ provide the source of lepton flavor violation.
The lepton flavor violating elements in $ \lambda_1 $
are then generated at $ M_W $ through the renormalization group evolution
with these couplings $ \lambda_2^{i4} $.
The renormalization group equations for the Yukawa couplings
are given in the model (1) by
\begin{eqnarray}
\mu \frac{d}{d\mu} \lambda_1^{aj}
= \frac{1}{(4\pi)^2} 
&& \hspace*{-5mm}
\left[
3 \sum_{k=1}^3 \sum_{b=1}^4 
\lambda_1^{ak} \lambda_1^{bk} \lambda_1^{bj}
+ \sum_{b=1}^4 \lambda_1^{bj} \lambda_2^{a4} \lambda_2^{b4}
\right.
\nonumber \\
&& \left.
+ \lambda_1^{aj} \left \{ 
\sum_{b=1}^4 \sum_{k=1}^3 ( \lambda_1^{bk} )^2
+ \lambda_3^2 + 3 \lambda_4^2
\right \}
- \lambda_1^{aj} (3 g_2^2 + 3 g_1^2)
\right] \ ,
\label{6}
\\
\mu \frac{d}{d\mu} \lambda_2^{a4} =
\frac{1}{(4\pi)^2} 
&& \hspace*{-5mm}
\left[
\lambda_2^{a4} \left \{ 
4 \sum_{b=1}^4 ( \lambda_2^{b4} )^2 + 2 \lambda_3^2 + 2 \lambda_4^2
\right \} \right. \nonumber \\
&& \left.
+ \sum_{b=1}^4 \sum_{i=1}^3 \lambda_2^{b4} \lambda_1^{ai} \lambda_1^{bi}
- \lambda_2^{a4} (3g_2^2 +g_1^2) \right] \ .
\end{eqnarray}
The renormalization group equations in the model (2)
are also obtained by setting $ \lambda_2^{44} = 0 $
in the above equations.
Then, we have the leading corrections for the off-diagonal elements
of the $ \lambda_1 $ coupling as
\begin{eqnarray}
\lambda_1^{aj} ( M_W ) - \lambda_1^{aj} ( M_{\rm G} )
\sim \frac{1}{( 4 \pi )^2} \lambda_1^{jj} \lambda_2^{j4} \lambda_2^{a4}
\log \frac{M_{\rm G}}{M_W} \ ( a \not= j ) \ .
\label{8}
\end{eqnarray}
Hence, even if $ \lambda_1^{aj} ( M_{\rm G} ) = 0 $ ($ a \not= j $)
at $ M_{\rm G} $ as given in eq.(\ref{3}), these flavor changing elements
of the $ \lambda_1 $ coupling arise at $ M_W $.
It should here be remarked that these corrections for $ \lambda_1 $
are in fact proportional to the corresponding diagonal elements
$ \lambda_1^{jj} $.
It should also be mentioned that in the model (2)
$ \lambda_1^{4j} \equiv 0 $ and $ \lambda_2^{44} \equiv 0 $,
which is ensured by the $ {\rm Z}_3 $ symmetry (\ref{4}).

The mass terms of the leptons are produced
by the vacuum expectation values (VEV's) of the Higgs fields,
\begin{eqnarray}
 \langle H_1 \rangle = 
 \left(
\begin{array}{c}
 v_1\\
 0\\
\end{array}
\right)\ ,\ 
\langle H_2 \rangle =
\left(
\begin{array}{c}
 0\\
 v_2\\
\end{array}
\right)\ ,\ 
\langle S \rangle = v_3 \ ,
\end{eqnarray}
where $ v = ( v_1^2 + v_2^2 )^{1/2} \simeq 174 {\rm GeV} $
and $ \tan \beta = v_2 / v_1 $.
These VEV's may acquire nonvanishing phases
introducing a new source of $CP$ violation.
We, however, do not consider this possibility for simplicity,
and therefore take the real VEV's.
Then, the charged lepton mass matrix is given at $ M_W $ by
\begin{eqnarray}
M_{\cal E} = \left( \begin{array}{cc}
\lambda_1^{ij} v_1 & \lambda_2^{i4} v_3 \\
\lambda_1^{4j} v_1 & \lambda_2^{44} v_3  \\
\end{array} \right) \ .
\label{10}
\end{eqnarray}
It should here be remembered that $ \lambda_2^{44} v_3 \rightarrow M_L $
in the model (2).  This $4 \times 4$ real matrix is diagonalized by
two orthogonal matrices $U_{\cal E}$ and $V_{\cal E}$ as
\begin{eqnarray}
 U_{\cal E}^{\rm T} M_{\cal E} V_{\cal E}
= {\rm diag.} ( m_e , m_\mu , m_\tau , M_E ) \ .
\end{eqnarray}
The neutral lepton mass matrix is also produced as
\begin{eqnarray}
M_{\cal N} = \left( \begin{array}{c}
- \lambda_2^{i4} v_3 \\ - \lambda_2^{44} v_3 \\
\end{array} \right) \ .
\end{eqnarray}
This $4 \times 1$ real matrix is diagonalized with
an orthogonal matrix $O_{\cal N}$ as
\begin{eqnarray}
O_{\cal N}^{\rm T} M_{\cal N} = \left( \begin{array}{c}
{\bf 0} \\ M_N \\ \end{array} \right) \ .
\label{13}
\end{eqnarray}
The masses of the ordinary charged leptons are given by
\begin{eqnarray}
m_{e_i} \simeq \lambda_1^{ii} v_1 \ ,
\end{eqnarray}
where the radiative corrections (\ref{8}) for the $ \lambda_1 $ coupling
are considered.  The masses of the exotic leptons are also given by
\begin{eqnarray}
M_E \simeq M_N = v_3 \
\left[ \sum_{a=1}^4 ( \lambda_2^{a4} )^2 \right]^{1/2} \ .
\end{eqnarray}
The zero modes in eq.(\ref{13})
are the three generations of ordinary neutrinos.
It is noticed that the orthogonal matrix $O_{\cal N}$
is not determined uniquely at this stage.
An arbitrary $ 3 \times 3 $ transformation, say $O_\nu^\prime$,
to redefine the (approximately) massless ordinary neutrinos
can be incorporated in $O_{\cal N}$.
The ordinary neutrinos may acquire very small Majorana masses
through the seesaw mechanism with the right-handed neutrinos
of ${\rm SU(2)_{\rm L}}$ singlet.
Then, the neutrino mixing matrix $O_{\cal N}$ is determined completely.

We here consider some relevant properties of the transformation matrices
$U_{\cal E}$, $V_{\cal E}$, and $O_{\cal N}$.
First, the orthogonal transformation $V_{\cal E}$,
which represents the right-handed charged lepton mixing,
diagonalizes the matrix $ M_{\cal E}^{\rm T} M_{\cal E} $.
By considering the structure of $ M_{\cal E}^{\rm T} M_{\cal E} $
from eq.(\ref{10}) with eq.(\ref{8}),
the right-handed ordinary-exotic lepton mixing
in $V_{\cal E}$, which is relevant for describing the gauge interactions
as seen below, is estimated in the leading order
(with $ M_E \sim v_3 $ for definiteness) as
\begin{eqnarray}
(V_{\cal E})_{i4} \simeq - (V_{\cal E})_{4i} \sim
( m_{e_i} / M_E ) \lambda_2^{i4} \ .
\label{16}
\end{eqnarray}
It should here be remarked that the right-handed lepton mixing
effects are suppressed sufficiently by the mass ratios $m_{e_i}/M_E$
with the experimental bound $M_E>97.0 \ {\rm GeV}$ \cite{l3},
even if the ordinary-exotic lepton couplings $ \lambda_2^{i4} $
take significant values.
On the other hand, the left-handed ordinary-exotic lepton mixings
in $U_{\cal E}$ and $O_{\cal N}$ are estimated as
\begin{eqnarray}
(U_{\cal E})_{i4} \simeq - (U_{\cal E})_{4i}
\simeq (O_{\cal N})_{i4} \simeq - (O_{\cal N})_{4j} ( O_\nu^\prime )_{ji}
\sim \lambda_2^{i4} \ ,
\end{eqnarray}
where the ambiguity in $ O_{\cal N} $ as mentioned above
is extracted explicitly by suitably choosing
the orthogonal transformation $ O_\nu^\prime $ of the ordinary neutrinos.
These left-handed mixings are no longer suppressed by $ m_{e_i} / M_E $.
Here, we note the similarity between $U_{\cal E}$ for the charged leptons
and $O_{\cal N}$ for the neutral leptons.
In fact, the effective combination of left-handed lepton mixing
is estimated as
\begin{eqnarray}
&&(U_{\cal E}^{\rm T} O_{\cal N})_{i4}
\simeq - (U_{\cal E}^{\rm T} O_{\cal N})_{4j} ( O_\nu^\prime )_{ji}
\sim ( m_{e_i} / M_E )^2 \lambda_2^{i4} \ ,
\label{18}
\\
&& (U_{\cal E}^{\rm T} O_{\cal N})_{ik} ( O_\nu^\prime )_{kj} - \delta_{ij}
\sim ( m_{e_i} / M_E ) ( m_{e_j} / M_E ) \lambda_2^{i4} \lambda_2^{j4} \ .
\label{19}
\end{eqnarray}
This similarity of the left-handed lepton mixings
described in terms of the $ \lambda_2^{i4} $ couplings
and the ordinary-exotic lepton mass ratios
is indeed stable against the radiative corrections.
It can be understood by noting the following facts.
(i) By setting $ \lambda_1^{ii} = 0 $ at $ M_{\rm G} $ in eq.(\ref{8}),
we have $ \lambda_1^{aj} \equiv 0 $ at any scales
as seen from eq.(\ref{6}).
Then, $ M_{\cal E} $ essentially coincides with $ M_{\cal N} $,
and they are diagonalized with the left-handed lepton transformation
$ U_{\cal E} = O_{\cal N} $ to eliminate
the $ \lambda_2^{i4} v_3 $ term.
(ii) The change $ \lambda_1 \rightarrow - \lambda_1 $
of the Yukawa coupling in $ M_{\cal E} $ is compensated
by the change $ e_i^c \rightarrow - e_i^c $
of the right-handed ordinary leptons.
This means that the sign change of $ \lambda_1 $
does not affect the left-handed lepton mixing.
Hence, by considering (i) and (ii)
we find that the difference between $ U_{\cal E} $ and $ O_{\cal N} $
should appear at the second order of $ \lambda_1^{aj} ( M_W )
\sim \lambda_1^{jj} \sim m_{e_j}/v_1 $.
(iii) By setting $ \lambda_2^{i4} ( M_{\rm G} ) = 0 $
so that $ \lambda_2^{i4} = 0 $ at any scales,
the $ \lambda_1 $ coupling remains to be diagonal
even if the radiative corrections are included.
Hence, in this limit with $ \lambda_2^{i4} = 0 $,
$ M_{\cal E} $ and $ M_{\cal N} $ are diagonal,
and the complete similarity follows
with $ U_{\cal E} = O_{\cal N} = {\bf 1}_4 $ ($ 4 \times 4 $ unit matrix).

It can be shown that the flavor violating gauge couplings
of leptons are suppressed naturally
due to the smallness of $(V_{\cal E})_{i4}$
and the similarity between $U_{\cal E}$ and $O_{\cal N}$.
The charged gauge interaction coupled to $W$ boson is given by
\begin{eqnarray}
{\cal L}_W
= -\frac{g_2}{\sqrt{2}} \ (U_{\cal E}^{\rm T} O_{\cal N})_{ab}
{\bar{\cal E}}_a \gamma^\mu P_{\rm L} {\cal N}_b W_\mu^-
+\frac{g_2}{\sqrt{2}} \ (V_{\cal E})_{4a} {\bar{\cal E}}_a \gamma^\mu
P_{\rm R} {\cal N}_4 W^-_\mu + {\rm h.c.} \ ,
\label{20}
\end{eqnarray}
where ${\cal E}_a \equiv (e_i,E)$ and ${\cal N}_a \equiv (\nu_i,N)$
($ a = 1 - 4 $) represent the four-component Dirac spinors
of the lepton mass eigenstates,
and $ P_{\rm L} $ and $ P_{\rm R} $ are the chiral projections
to the left-handed and right-handed parts, respectively.
This charged gauge interaction is relevant
for the $ \mu \rightarrow e \gamma $ decay.
The lepton flavor violating terms
with the factors $ (U_{\cal E}^{\rm T} O_{\cal N})_{ab} $ ($a \neq b$)
are apparently induced in the left-handed couplings of eq.(\ref{20}).
However, due to the similarity between $U_{\cal E}$ and $O_{\cal N}$
as shown in eqs.(\ref{18}) and (\ref{19}),
they are almost rotated out with the suitable redefinition
$ O_\nu^\prime $ of the ordinary neutrinos neglecting
the tiny neutrino masses $ m_{\nu_i} $.
The flavor changing right-handed couplings
are also suppressed by the small factors
$(V_{\cal E})_{4i}$, as shown in eq.(\ref{16}).
Hence, the effects of the charged gauge interaction
appear to be extremely small for the lepton flavor violating processes.

The neutral gauge interaction of charged leptons
coupled to $Z$ boson is given by
\begin{eqnarray}
{\cal L}_Z = {\cal L}_Z^0
+ \frac{1}{2}
\left( g_2 \cos \theta_W + g_1 \sin \theta_W \right)
(V_{\cal E})_{4a} (V_{\cal E})_{4b}
{\bar{\cal E}_a} \gamma^\mu P_{\rm R} {\cal E}_b Z_\mu \ ,
\label{21}
\end{eqnarray}
where $ {\cal L}_Z^0 $ represents the usual flavor diagonal part
which is obtained by turning off the ordinary-exotic lepton mixing.
The lepton flavor violating term appears
only in the right-handed couplings of eq.(\ref{21}).
This is because the left-handed leptons all form
the $ {\rm SU(2)_{\rm L}} $ doublets.
Since the FCNC's of ordinary leptons in eq.(\ref{21}) are proportional to
$(V_{\cal E})_{4i} (V_{\cal E})_{4j}$,
their effects are fairly suppressed
by the small mass ratios $ (m_{e_i}/M_E)(m_{e_j}/M_E) $.
If the quark singlets $ ( D^c , D ) $ are also present,
they may have the flavor violating coupling
which has the common ${\rm E_6}$ origin with the $ \lambda_2 $ coupling
of $ ( L , L^c ) $.
In this case, the FCNC's of ordinary quarks appear as well
to be negligibly small due to the suppression factors
$ (m_{d_i}/M_D)(m_{d_j}/M_D) $.

We now estimate the contributions to the $ \mu \rightarrow e \gamma $ decay
which are provided by these gauge interactions.
We will in fact observe below that these contributions
are negligibly small.
The decay amplitude of $ \mu \rightarrow e \gamma $
is generally given by
\begin{eqnarray}
T ( \mu \rightarrow e \gamma ) = e \epsilon^{\alpha *}
{\bar u}_e ( p - q ) \left[ i \sigma_{\alpha \beta} q^\beta
( A_{\rm L} P_{\rm L} + A_{\rm R} P_{\rm R} ) \right] u_\mu (p) \ .
\end{eqnarray}
Then, the decay rate is calculated by
\begin{eqnarray}
\Gamma ( \mu \rightarrow e \gamma )
= \frac{1}{16 \pi} e^2 m_\mu^3 ( | A_{\rm L} |^2 + | A_{\rm R} |^2 ) \ .
\end{eqnarray}
The contributions to the amplitudes $ A_{\rm L} $ and $ A_{\rm R} $
are provided by the one-loop diagrams
with the intermediate states of the neural leptons
($ W $ mediated) and charged leptons ($ Z $ mediated).
We can see from eqs.(\ref{16}) -- (\ref{19})
that all these diagrams include the significant suppression factor
$ (m_e/M_E)(m_\mu/M_E) \sim 10^{-8} $.
Then, the $ W $ and $ Z $ contributions to the decay amplitude
are given by
\begin{eqnarray}
A^{(W)}_{\rm L,R} \sim A^{(Z)}_{\rm L,R}
\sim \frac{1}{32 \pi^2} \frac{m_e m_\mu^2}{M_W^2 M_E^2} \
\lambda_2^{14} \lambda_2^{24} \ .
\label{24}
\end{eqnarray}
Therefore, the flavor violating gauge couplings
induced by the ordinary-exotic lepton mixing
provide tiny contributions $ \sim 10^{-21} $
to the branching ratio ${\rm Br}( \mu \rightarrow e \gamma ) $.
This feature is actually confirmed by numerical calculations.

\section{The slepton mixing}
In the supersymmetric model, significant effects
of flavor violation may be obtained through the slepton mixing.
It is expected that the supersymmetry breaking is provided
in the minimal supergravity model
with the soft terms at the gravitational scale
$ M_{\rm G} \sim 10^{18} \ {\rm GeV} $,
including the common slepton masses squared
$ {m^2_{\tilde l}}_{ab} = m_0^2 \delta_{ab} $, etc.,
the $ A $ terms $ A_k = a_0 m_0 \lambda_k $,
and the gaugino masses $ m_1 $ and $ m_2 $.
We have seen that in the present sort of models
the original source of lepton flavor violation
is the ordinary-exotic lepton couplings $ \lambda_2^{i4} $.
At the tree-level with the universal soft supersymmetry breaking terms,
however, the flavor violating effects of these couplings
are still suppressed by the ordinary-exotic lepton mass ratios
$ m_{e_i} / M_E $,
since the lepton and slepton mass matrices are diagonalized
simultaneously in this limit.
Significant effects of flavor violation are rather provided
from the radiative corrections on the soft terms
which are generated with the ordinary-exotic lepton couplings
$ \lambda_2^{i4} $.

The soft mass terms and $ A $-terms at the electroweak scale $ M_W $
deviate from the universal forms through
the renormalization group evolution \cite{Hall:1986dx}.
In the leading order approximation the corrections
are given for the model (1) by
\begin{eqnarray}
\Delta {m^2_{\tilde l}}_{aa}
&=& - \frac{1}{(4\pi)^2}
\left[ 2 m_0^2 (3+a_0^2)
\left\{ ( \lambda_1^{aa} )^2 + ( \lambda_2^{a4} )^2 \right\}
- (6m_2^2 g_2^2 +2 m_1^2 g_1^2) \right]
\log \frac{M_{\rm G}}{M_W} \ ,
\\
\Delta {m^2_{\tilde l}}_{ab}
&=& -\frac{2}{(4\pi)^2} m_0^2 (3+a_0^2)
\lambda_2^{a4} \lambda_2^{b4} \log \frac{M_{\rm G}}{M_W} \ \ (a \neq b) \ ,
\\
\Delta {m^2_{{\tilde e}^c}}_{ij}
&=& -\frac{\delta_{ij}}{(4\pi)^2}
\left[ 4 m_0^2 ( \lambda_1^{ii} )^2 (3+a_0^2) - 8 g_1^2 m_1^2 \right]
\log \frac{M_{\rm G}}{M_W} \ ,
\\
\Delta m^2_{{\tilde L}^c}
&=& - \frac{1}{(4\pi)^2}
\left[ 2 m_0^2 (3+a_0^2) \sum_{a=1}^4 ( \lambda_2^{a4} )^2
-(6 g_2^2 m_2^2 +2 g_1^2 m_1^2) \right]
\log \frac{M_{\rm G}}{M_W} \ ,
\\
\Delta A_1^{ii}
&=&
-\frac{1}{(4\pi)^2} \lambda_1^{ii}
\left[ {\rule[-3mm]{0mm}{10mm}\ }
3 a_0 m_0 \left\{ ( \lambda_2^{i4} )^2 + \lambda_3^2 + 3 \lambda_4^2
+ 3 ( \lambda_1^{ii} )^2 + \sum_{j=1}^3 ( \lambda_1^{jj} )^2
\right\} \right.
\nonumber \\
&& \left. - 3 a_0 m_0 ( g_2^2 + g_1^2 )
- 6 ( g_2^2 m_2 + g_1^2 m_1)
{\rule[-3mm]{0mm}{10mm}\ } \right]
\log \frac{M_{\rm G}}{M_W} \ ,
\\
\Delta A_1^{aj}
&=& -\frac{3}{(4\pi)^2} a_0 m_0 \lambda_1^{jj}
\lambda_2^{a4} \lambda_2^{j4} \log \frac{M_{\rm G}}{M_W} \ \ (a \neq j) \ ,
\\
\Delta A_2^{a4}
&=& - \frac{1}{(4\pi)^2} \lambda_2^{a4}
\left[ {\rule[-3mm]{0mm}{10mm}\ }
3 a_0 m_0 \left\{
4 \sum_{b=1}^4 ( \lambda_2^{b4} )^2 + 2 \lambda_3^2 + 2 \lambda_4^2
+ ( \lambda_1^{aa} )^2 \right\} \right.
\nonumber \\
&& \left. - a_0 m_0 (3 g_2^2 +g_1^2)
- 2 (3 g_2^2 m_2 + g_1^2 m_1) 
{\rule[-3mm]{0mm}{10mm}\ } \right]
\log \frac{M_{\rm G}}{M_W} \ .
\end{eqnarray}
We should set $ \lambda_2^{44} = 0 $ for the model (2)
in these formulas.
Because of these radiative corrections,
the slepton mass matrices and the corresponding lepton mass matrices
are no longer diagonalized simultaneously,
providing the new source of lepton flavor violation
in the slepton mixing.
Then, this slepton mixing is expected to contribute significantly
to the $\mu \rightarrow e \gamma$ decay.
The lepton-slepton interactions with the neutralinos
$ \chi^0_\alpha $ ($ \alpha = 1 - 5 $)
and the charginos $ \chi^\pm_\kappa $ ($ \kappa = 1, 2 $),
which are relevant for $\mu \rightarrow e \gamma$,
are given, respectively, by
\begin{eqnarray}
{\cal L}_{\chi^0} =
N^{\rm R}_{a \alpha A} {\bar{\cal E}}_a
P_{\rm R} \chi^0_\alpha {\tilde{\cal E}}_A
+ N^{\rm L}_{a \alpha A} {\bar{\cal E}}_a
P_{\rm L} \chi^0_\alpha {\tilde{\cal E}}_A
+ \ {\rm h.c.}
\end{eqnarray}
and
\begin{eqnarray}
{\cal L}_{\chi^\pm} = C^{\rm R}_{a \kappa K} {\bar{\cal E}}_a
P_{\rm R} \chi^-_\kappa {\tilde{\cal N}}_K
+ C^{\rm L}_{a \kappa K} {\bar{\cal E}}_a
P_{\rm L} \chi^-_\kappa {\tilde{\cal N}}_K
+ \ {\rm h.c.} \ ,
\end{eqnarray}
where $ {\tilde{\cal E}}_A $ ($ A = 1 - 8 $)
and $ {\tilde{\cal N}}_K $ ($ K = 1 - 5 $) are the slepton mass eigenstates.
The coupling coefficients are given in terms of the transformation matrices
to diagonalize the mass matrices of
leptons, sleptons, charginos, and neutralinos.

We here evaluate the decay rate of $ \mu \rightarrow e \gamma $
which is provided by the slepton mixing.
By calculating the one-loop diagrams
for the $\mu \rightarrow e \gamma$ process
mediated by the neutralinos and charginos, as shown in fig.\ref{meg},
we obtain the contributions to the decay amplitudes
\cite{hisano}
\begin{eqnarray}
A^{({\rm n})}_{\rm R} &=& \frac{1}{32 \pi^2} \sum_{A, \alpha} \left[
\frac{m_\mu}{M^2_{{\tilde{\cal E}}_A}}
N^{\rm R}_{1 \alpha A} N^{R *}_{2 \alpha A} f_1 (x_{\alpha A})
+ \frac{M_{\chi^0_\alpha}}{M^2_{{\tilde{\cal E}}_A}}
N^{\rm R}_{1 \alpha A} N^{L *}_{2 \alpha A} f_2 (x_{\alpha A}) \right] \ ,
\\
A^{({\rm n})}_{\rm L} &=& A^{({\rm n})}_{\rm R} ( R \leftrightarrow L ) \ ,
\\
A^{({\rm c})}_{\rm R} &=& \frac{1}{32 \pi^2} \sum_{K, \kappa} \left[
\frac{m_\mu}{M^2_{{\tilde{\cal N}}_K}}
C^{\rm R}_{1 \kappa K} C^{R *}_{2 \kappa K} f_3 (x_{\kappa K})
+ \frac{M_{\chi^-_\kappa}}{M^2_{{\tilde{\cal N}}_K}}
C^{\rm R}_{1 \kappa K} C^{L *}_{2 \kappa K} f_4 (x_{\kappa K}) \right] \ ,
\\
A^{({\rm c})}_{\rm L} &=& A^{({\rm c})}_{\rm R} ( R \leftrightarrow L ) \ ,
\end{eqnarray}
where $ f_1 - f_4 $ are certain functions
of $ x_{\alpha A}
\equiv ( M_{\chi^0_\alpha}/M_{{\tilde{\cal E}}_A} )^2 $
and $ x_{\kappa K}
\equiv ( M_{\chi^-_\kappa}/M_{{\tilde{\cal N}}_K} )^2 $.
We estimate below $ {\rm Br} ( \mu \rightarrow e \gamma ) $
with these contributions for the model (1) and (2).
We will see that these models provide quite different results.
In fact, these supersymmetric contributions are rather suppressed
in the model (1) with the extra factor involving
$ m_e $ as well as $ m_\mu $.

\subsection{Model (1)}

The charged slepton mass matrix is given as
\begin{eqnarray}
{\cal M}^2_{\tilde {\cal E}} = \left( \begin{array}{cc}
M^2_{{\tilde{\cal E}}_{{\rm L} {\rm L}}}
& M^2_{{\tilde{\cal E}}_{{\rm L} {\rm R}}} \\
M^2_{{\tilde{\cal E}}_{{\rm R} {\rm L}}}
& M^2_{{\tilde{\cal E}}_{{\rm R} {\rm R}}} \end{array} \right) \ ,
\end{eqnarray}
where $ {\rm L} $ and $ {\rm R} $ represent the chirality
of the corresponding leptons.
The neutral slepton mass matrix is also given in a similar form.
In order to see the flavor changing structure
of the lepton-slepton-gaugino interactions,
it is suitable to transform this slepton mass matrix
by the orthogonal matrices which diagonalize the lepton mass matrix:
\begin{eqnarray}
{{\cal M}^2_{\tilde {\cal E}}}^\prime
= \left( \begin{array}{cc} U_{\cal E}^{\rm T} & 0 \\
0 & V_{\cal E}^{\rm T} \end{array} \right)
{\cal M}^2_{\tilde {\cal E}}
\left( \begin{array}{cc} U_{\cal E} & 0 \\
0 & V_{\cal E} \end{array} \right) \ .
\label{39}
\end{eqnarray}
The universality for the soft supersymmetry breaking at $ M_{\rm G} $
is violated at $ M_W $ due to the renormalization group effects.
Then, the reduced slepton mass matrix
$ {{\cal M}^2_{\tilde {\cal E}}}^\prime $
involves the flavor changing elements.

In the model (1), the components of the slepton mass matrix
are given including the renormalization group effects as
\begin{eqnarray}
M^2_{{\tilde{\cal E}}_{{\rm L}{\rm L}}}
&=& x_{{\rm L}{\rm L}} M_{\cal E} M_{\cal E}^{\rm T}
+ y_{{\rm L}{\rm L}} m_0^2 {\bf 1}_4
+ m_0^2 \left( \begin{array}{cc}
c_{{\rm L}{\rm L}}^{ij} ( \lambda_1^{ii} )^2
+ c_{{\rm L}{\rm L}}^{ji} ( \lambda_1^{jj} )^2
  & d_{{\rm L}{\rm L}}^i ( \lambda_1^{ii} )^2 \\
d_{{\rm L}{\rm L}}^j ( \lambda_1^{jj} )^2 & 0
\end{array} \right) \ , \\
M^2_{{\tilde{\cal E}}_{{\rm L}{\rm R}}}
&=& ( M^2_{{\tilde{\cal E}}_{{\rm R}{\rm L}}} )^{\rm T}
\nonumber \\
&=& x_{{\rm L}{\rm R}} m_0 M_{\cal E}
+ x_{{\rm L}{\rm R}}^\prime m_0
\left( \begin{array}{cc} {\bf 0} & M_{\cal N} \end{array} \right)
+ m_0^2 \left( \begin{array}{cc}
\delta_{ij} f_{{\rm L}{\rm R}}^i \lambda_1^{ii}
& g_{{\rm L}{\rm R}}^i ( \lambda_1^{ii} )^2 \\
{\bf 0} & 0 \end{array} \right) \ , \\
M^2_{{\tilde{\cal E}}_{{\rm R} {\rm R}}}
&=& x_{{\rm R}{\rm R}} M_{\cal E}^{\rm T} M_{\cal E}
+ y_{{\rm R}{\rm R}} m_0^2 {\bf 1}_4
+ z_{{\rm R}{\rm R}} m_0^2
\left( \begin{array}{cc} {\bf 0} & {\bf 0} \\
{\bf 0} & 1 \end{array} \right)
+ m_0^2 \left( \begin{array}{cc}
\delta_{ij} c_{{\rm R}{\rm R}}^i ( \lambda_1^{ii} )^2 & {\bf 0} \\
{\bf 0} & 0 \end{array} \right) \ ,
\label{42}
\end{eqnarray}
where $ x_{{\rm L}{\rm L}} , ... \sim 1 $ are the relevant parameters.
The terms proportional to the $ 4 \times 4 $ unit matrix $ {\bf 1}_4 $
as well as those given by the charged lepton mass matrix $ M_{\cal E} $
are diagonalized by the transformation in eq.(\ref{39}).
The term given by the neutral lepton mass matrix $ M_{\cal N} $
is almost diagonalized by the similarity of
$ U_{\cal E}^{\rm T} O_{\cal N} $.
The third term of eq.(\ref{42}) provides
a contribution $ ( V_{\cal E} )_{4i} ( V_{\cal E} )_{4j} $
for the slepton mixing, which is substantially
suppressed by the lepton mass ratios
$ ( m_{e_i} / M_E ) ( m_{e_j} / M_E ) $.
The remaining terms involve the Yukawa couplings
$ \lambda_1^{ii} \sim m_{e_i}/v_1 $.
Among these contributions in the model (1),
the significant flavor mixing arises from the fourth term
of eq.(\ref{42}) in the right-handed charged slepton sector as
\begin{eqnarray}
( M^{2 \prime}_{{\tilde{\cal E}}_{{\rm R}{\rm R}}} )_{ij} / m_0^2
\sim ( V_{\cal E}^{\rm T} )_{ii} ( \lambda_1^{ii} )^2 ( V_{\cal E} )_{ij}
+ ( V_{\cal E}^{\rm T} )_{ij} ( \lambda_1^{jj} )^2 
( V_{\cal E} )_{jj} \ .
\label{43}
\end{eqnarray}
It is here noted that the right-handed ordinary charged lepton mixing
induced by the ordinary-exotic lepton mixing
is related to the charged lepton masses as
\begin{eqnarray}
( V_{\cal E} )_{ij} \sim \frac{m_{e_i} m_{e_j}}{m_{e_i}^2 + m_{e_j}^2}
\lambda_2^{i4} \lambda_2^{j4} \ ( i \not= j ) \ .
\label{44}
\end{eqnarray}

By considering these arguments on the slepton mixing,
the dominant contribution from the neutralino couplings
to the decay amplitude of $ \mu \rightarrow e \gamma $ is estimated as
\begin{eqnarray}
A^{({\rm n})}_{\rm L}
\sim \frac{1}{32 \pi^2}
\frac{\lambda_3 \lambda_1^{22} v_2 v_3 }{m_0^3} \ ( \lambda_1^{22} )^2
( m_e / m_\mu ) \lambda_2^{14} \lambda_2^{24} \ .
\label{45}
\end{eqnarray}
Here, the right-handed $ {\tilde \mu}^c $-$ {\tilde e}^c $
slepton mixing $ \sim ( \lambda_1^{22} )^2 ( V_{\cal E} )_{12} $
given by eqs.(\ref{43}) and (\ref{44}) is used
for the $ e_{\rm R} $-$ {\tilde \mu}^c $-$ \chi^0 $ vertex,
and the chirality flip $ \sim \lambda_3 \lambda_1^{22} v_2 v_3 $
of the intermediate $ {\tilde \mu} $ state
is provided by the $ | F_{H_1} |^2 $ term.
The chargino couplings provide a contribution
of the same order as eq.(\ref{45}),
where the chirality flipping $ {\tilde N} $-$ {\tilde e}^c $
slepton mixing given by
$ ( O_{\cal E}^{\rm T} M^2_{{\tilde{\cal N}}_{{\rm L}{\rm R}}} )_{41} / m_0
\sim ( m_e / m_0 ) \tan \beta \lambda_2^{14} $
is used for the $ e_{\rm R} $-$ {\tilde N} $-$ \chi_0 $ vertex,
and the the left-handed $ {\tilde N} $-$ {\tilde \nu}_\mu $
slepton mixing in the intermediate state
is provided by
$ ( O_{\cal E}^{\rm T} M^2_{{\tilde{\cal N}}_{{\rm L}{\rm L}}} )_{42} / m_0^2
\sim ( m_\mu / M_W )^2 \tan \beta \lambda_2^{24} $.
Accordingly, in the model (1) the contributions to the decay amplitude
are dominantly given by
\begin{eqnarray}
A^{({\rm n})}_{\rm L} \sim A^{({\rm c})}_{\rm L}
\sim \frac{1}{32 \pi^2} \tan^3 \beta \frac{m_e m_\mu^2}{M_W^2 m_0^2}
\lambda_2^{14} \lambda_2^{24} \ ,
\label{46}
\end{eqnarray}
where $ v_3 \sim m_0 $ is considered for definiteness.
These contributions should be compared to the non-supersymmetric
contributions given in eq.(\ref{24}).
Although these supersymmetric contributions have the specific dependence
on the relevant lepton masses similar to eq.(\ref{24}),
it can be enhanced significantly by the power of $ \tan \beta $.
We roughly estimate the branching ratio as
\begin{eqnarray}
{\rm Br} ( \mu \rightarrow e \gamma ) & \sim &
\frac{3 \alpha}{32 \pi} \tan^6 \beta
\left( \frac{m_e m_\mu}{m_0^2} \right)^2
\left( \lambda_2^{14} \lambda_2^{24} \right)^2
\nonumber \\
& \sim & 10^{-13} \
( \tan \beta \sim 20 \ , \
 \lambda_2^{14} \sim \lambda_2^{24} \sim 0.3 ) \ .
\label{47}
\end{eqnarray}
It should here be remarked as seen in eq.(\ref{46})
that in the model (1) the chirality of the charged leptons
is mainly specified as $ \mu^-_{\rm L} \rightarrow e^-_{\rm R} \gamma $.

The dependence of the decay amplitudes on both $ m_e $ and $ m_\mu $,
as seen in eq.(\ref{46}), can be understood as follows
by means of approximate flavor symmetries.
Set $ \lambda_1^{11} ( M_{\rm G} ) = 0 $ in eq.(\ref{3}).
Then, we can make $ \lambda_2^{14} ( M_{\rm G} ) = 0 $
by a suitable transformation of the left-handed lepton doublets
$ l_1 $ and $ l_4 = L $, while keeping $ \lambda_1 ( M_{\rm G} ) $
with $ \lambda_1^{11} ( M_{\rm G} ) = 0 $.
In this limit, a flavor symmetry $ {\rm U(1)}_e $
appears under the phase transformation
$ l_1 \rightarrow {\rm e}^{i \alpha} l_1 $
and $ e^c_1 \rightarrow {\rm e}^{-i \alpha} e^c_1 $.
If the universal form is assumed for the supersymmetry breaking
terms at $ M_{\rm G} $, this flavor symmetry $ {\rm U(1)}_e $
is still preserved in the limit $ \lambda_1^{11} ( M_{\rm G} ) = 0 $.
This is the case even if the renormalization group corrections
are included.
Therefore, we find that in the limit  $ \lambda_1^{11} ( M_{\rm G} ) = 0 $
corresponding to $ m_e = 0 $, the $ \mu \rightarrow e \gamma $ decay
is prevented for the electron number conservation by the $ {\rm U(1)}_e $.
Similarly, the decay amplitudes of $ \mu \rightarrow e \gamma $
depend on $ m_\mu $ as well
due to the approximate $ {\rm U(1)}_\mu $ symmetry.
It should further be remarked that even for the case
with several pairs of vectorlike lepton doublets,
the $ \mu \rightarrow e \gamma $ decay amplitudes are dominantly
given by eq.(\ref{46}) in the model (1)
without the bare mass term $ M_L L L^c $.
This can be justified by extending readily the above symmetry argument.

In order to confirm the estimate
of $ {\rm Br} ( \mu \rightarrow e \gamma ) $ given in eq.(\ref{47}),
we have made numerical calculations
by taking reasonable values of the model parameters.
In the model (1), the most important parameters are $ \tan \beta $
and $ \lambda_2^{i4} $ for the ordinary-exotic lepton mixing.
The values of $ \lambda_1^{ii}$ are determined
so as to reproduce the actual lepton masses $m_e,\ m_\mu,\ m_\tau$.
The exotic leptons $ E $ and $ N $ have typically
the masses $ M_E \simeq M_N \sim 100 {\rm GeV} $
by taking $ \lambda_2^{44} \sim 0.3 $ and $ v_3 \sim 500 {\rm GeV} $.

\begin{figure}
\begin{center}
\includegraphics[height=9cm]{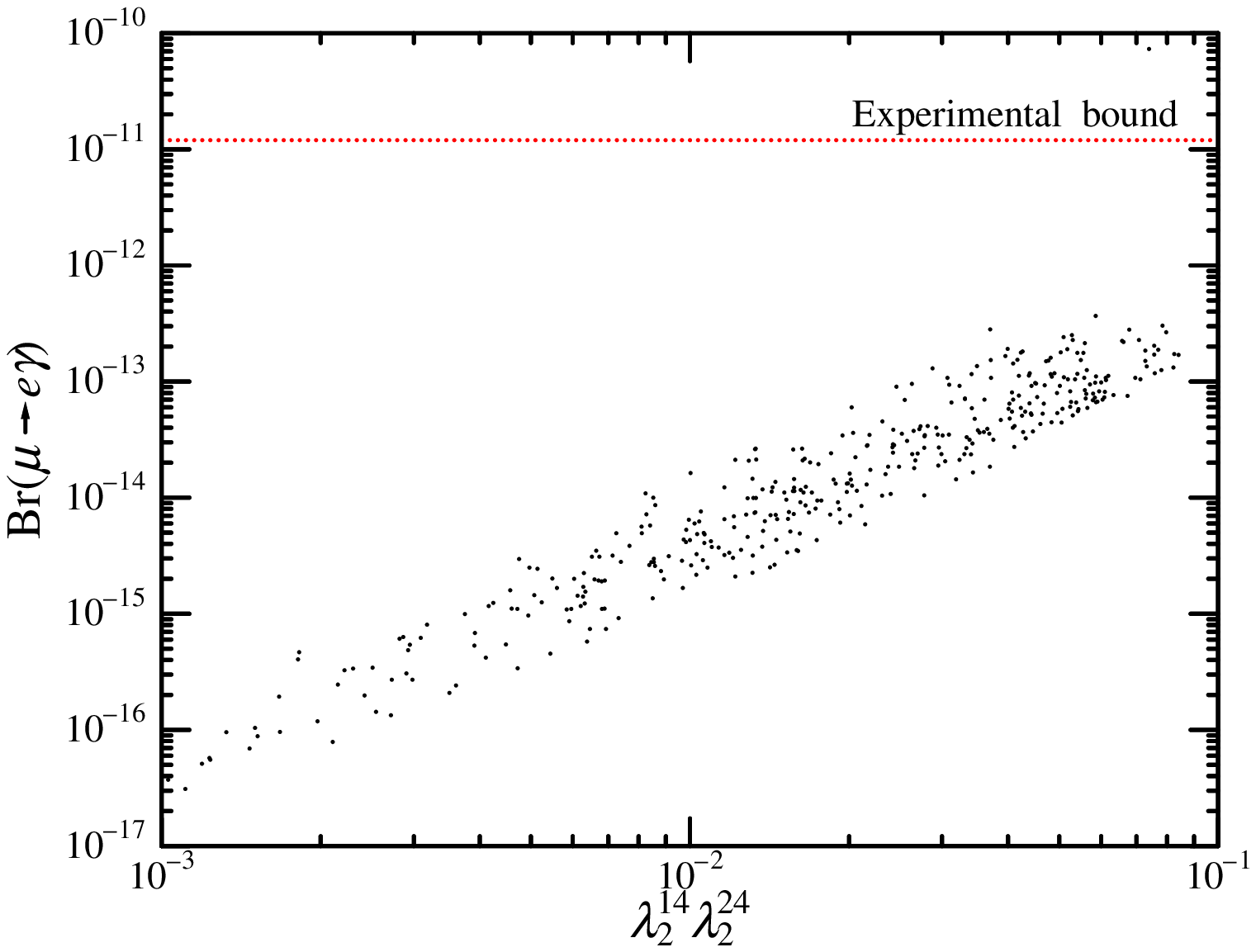}
\end{center}
\caption{
The relation between Br($\mu \rightarrow e \gamma$)
and $ \lambda_2^{14} \lambda_2^{24} $ is plotted in the model (1)
for $ \tan \beta = 20 $.
The points represent the branching ratio
for random values of $ \lambda_2^{i4} $.
\label{l2v1}}
\end{figure}

In fig.\ref{l2v1}, Br($\mu \rightarrow e \gamma$) is shown
as a function of $ \lambda_2^{14} \lambda_2^{24} $
by taking randomly the values of $ \lambda_2^{i4} $.
We have also taken typically $ \tan \beta = 20 $,
$ v_3 = 500 {\rm GeV} $, $ m_0 = 100 {\rm GeV} $, $ a_0 = 1 $,
$ m_1 = 100 {\rm GeV} $, $ \lambda_2^{44} = 0.3 $,
$ \lambda_3 = 0.6 $ and $ \lambda_4 = 0.5 $.
It is clearly observed in fig.\ref{l2v1}
that Br($\mu \rightarrow e \gamma$) is roughly proportional
to $ ( \lambda_2^{14} \lambda_2^{24} )^2 $,
as indicated in eq.(\ref{47}).

\begin{figure}
\begin{center}
\includegraphics[height=9cm]{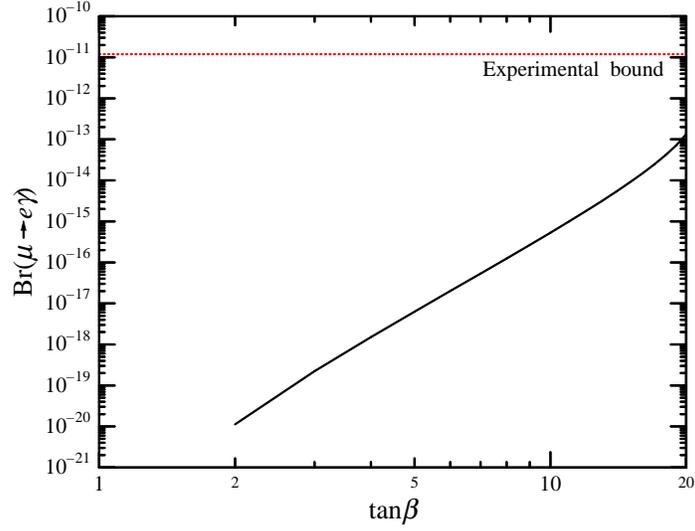}  
\end{center}
\caption{
The relation between Br($\mu \rightarrow e \gamma$)
and $\tan \beta$ is plotted in the model (1)
for $ \lambda_2^{a4} = 0.3 $.
\label{tanv1}}
\end{figure}

The relation between Br($\mu \rightarrow e \gamma$) and $\tan \beta$
is shown in fig.\ref{tanv1}.
Here, we have taken typically
$ v_3 = 500 {\rm GeV} $, $ m_0 = 200 {\rm GeV} $, $ a_0 = 0 $,
$ m_1 = 200 {\rm GeV} $, $ \lambda_2^{a4} = 0.3 $,
$ \lambda_3 = 0.6 $ and $ \lambda_4 = 0.5 $.
This result indicates the strong $\tan \beta$ dependence
as $ {\rm Br} ( \mu \rightarrow e \gamma ) \propto \tan^6 \beta $.
It is, in particular, interesting that if $ \tan \beta $
is larger than 20, $ {\rm Br} ( \mu \rightarrow e \gamma ) $
might be comparable to the experimental bound
$ 1.2 \times 10^{-11} $ \cite{brbound}.

As for the $ \mu \rightarrow 3e $ decay,
the penguin diagrams associated with fig.\ref{meg}
provide dominant contributions in the present model
with $ {\rm Br}( \mu \rightarrow 3e )/{\rm Br}( \mu \rightarrow e \gamma )
\sim 7 \times 10^{-3} $, as is usually the case \cite{hisano}.
The tree-level $ Z $ mediated FCNC's given in eq.(\ref{21})
provide only a contribution smaller by a few orders.
We have also the supersymmetric contributions
to $ \tau \rightarrow e \gamma $ and 
$ \tau \rightarrow \mu \gamma $, which are estimated in eq.(\ref{47})
by replacing the relevant charged lepton masses
and the $ \lambda_2^{i4} $ couplings.
In particular, $ {\rm Br} ( \tau \rightarrow \mu \gamma ) $
is enhanced as
\begin{eqnarray}
{\rm Br} ( \tau \rightarrow \mu \gamma )
& \sim & \left( \frac{m_\mu m_\tau}{m_e m_\mu} \right)^2
{\rm Br} ( \mu \rightarrow e \gamma )
\nonumber \\
& \sim & 10^{-7} \
( \tan \beta \sim 20 \ , \
\lambda_2^{24} \sim \lambda_2^{34} \sim 0.3 ) \ ,
\label{48}
\end{eqnarray}
which can be comparable to the experimental bound
$ 1.1 \times 10^{-6} $ \cite{Ahmed:1999gh}.
Therefore, in the model (1) the $ \tau \rightarrow \mu \gamma $ decay
seems to be more promising to observe the lepton flavor violation
due to the ordinary-exotic lepton mixing with supersymmetry.

The $ \lambda_2 $ coupling may also contribute
to the flavor violation in the quark sector
if the quark singlets $ ( D^c , D ) $ survive as well
to the electroweak scale in the ${\rm E_6}$ unification.
Then, we have dominantly the right-handed squark mixings such as
$ ( M^{2 \prime}_{{\tilde{\cal D}}_{{\rm R}{\rm R}}} )_{12} / m_0^2
\sim ( m_s m_d / v^2 ) \lambda_2^{14} \lambda_2^{24} \tan^2 \beta
\sim 10^{-8} \lambda_2^{14} \lambda_2^{24} \tan^2 \beta $,
which are the same as the right-handed slepton mixings
given in eqs.(\ref{43}) with (\ref{44}).
It is clearly found that even for $ \tan \beta \sim 50 $
this squark mixing is much smaller than the bound $ \sim 10^{-2} $
obtained from the $ K^0 $-$ {\bar K}^0 $ system \cite{Gabbiani:1996}.

\subsection{Model (2)}

In the model (2), we have more significant contributions
for the right-handed decay amplitudes.
This is because the $ {\rm U(1)}_e \times {\rm U(1)}_\mu $
considered in the case of model (1) is substantially
violated due to the presence of bare mass term $ M_L LL^c $
of the vectorlike lepton doublets.
Actually, the soft supersymmetry breaking mass terms
for the ordinary slepton doublets $ {\tilde l} $
and the vectorlike slepton doublet $ {\tilde L} $
acquire the renormalization group corrections
in somewhat different way with $ \lambda_2^{44} \equiv 0 $.
Then, we have an extra terms for $ M^2_{{\tilde{\cal E}}_{{\rm L}{\rm L}}} $
and $ M^2_{{\tilde{\cal N}}_{{\rm L}{\rm L}}} $ as
\begin{eqnarray}
z_{{\rm L}{\rm L}} m_0^2
\left( \begin{array}{cc} {\bf 0} & {\bf 0} \\
{\bf 0} & 1 \end{array} \right)
\rightarrow
M^2_{{\tilde{\cal E}}_{{\rm L}{\rm L}}} \ , \
M^2_{{\tilde{\cal N}}_{{\rm L}{\rm L}}} \ .
\end{eqnarray}
This extra term provides the significant mixing
for the left-handed sleptons as
\begin{eqnarray}
( M^{2 \prime}_{{\tilde{\cal E}}_{{\rm L}{\rm L}}} )_{ij} / m_0^2
\sim ( M^{2 \prime}_{{\tilde{\cal N}}_{{\rm L}{\rm L}}} )_{ij} / m_0^2
\sim ( U_{\cal E}^{\rm T} )_{i4} ( U_{\cal E} )_{4j}
\sim \lambda_2^{i4} \lambda_2^{j4} \ ,
\label{50}
\end{eqnarray}
where the similarity $ U_{\cal E} \simeq O_{\cal N} $ is considered.
These left-handed slepton mixings can be taken
for the one-loop diagrams similar to those considered
in the model (1) with chirality change $ {\rm L} \leftrightarrow {\rm R} $.
Then, we obtain the contributions
to the $ \mu \rightarrow e \gamma $ decay amplitude as
\begin{eqnarray}
A^{({\rm n})}_{\rm R} \sim A^{({\rm c})}_{\rm R}
\sim \frac{1}{32 \pi^2} \tan \beta \frac{m_\mu}{m_0^2}
\lambda_2^{14} \lambda_2^{24} \ .
\label{51}
\end{eqnarray}
It is here important that these supersymmetric contributions
are no longer suppressed by the very small extra factor
$ m_e m_\mu / M_W^2 $ as seen in eq.(\ref{46}) for the model (1).
We roughly estimate the branching ratio as
\begin{eqnarray}
{\rm Br} ( \mu \rightarrow e \gamma ) & \sim &
\frac{3 \alpha}{32 \pi} \tan^2 \beta
\left( \frac{M_W}{m_0} \right)^4
\left( \lambda_2^{14} \lambda_2^{24} \right)^2
\nonumber \\
& \sim & 10^{-11} \
( \tan \beta \sim 1 \ , \
 \lambda_2^{14} \sim \lambda_2^{24} \sim 0.01 ) \ .
\label{52}
\end{eqnarray}
As seen in eq.(\ref{51}), the chirality of the charged leptons
is mainly specified as $ \mu^-_{\rm R} \rightarrow e^-_{\rm L} \gamma $
in the model (2) contrary to the case of model (1).
It is also mentioned that this significant supersymmetric contribution
to $ {\rm Br} ( \mu \rightarrow e \gamma ) $
may be obtained even in the model (1),
if the soft slepton masses squared $ m_{\tilde l}^2 $
are different between $ {\tilde l}_i $ and $ {\tilde l}_4 = {\tilde L} $
already at $ M_{\rm G} $ for some reason.

We have also made numerical calculations
to evaluate $ {\rm Br} ( \mu \rightarrow e \gamma ) $
in the model (2) with the bare mass term,
which is expected to be much larger than in the model (1).

\begin{figure}
\begin{center}
\includegraphics[height=9cm]{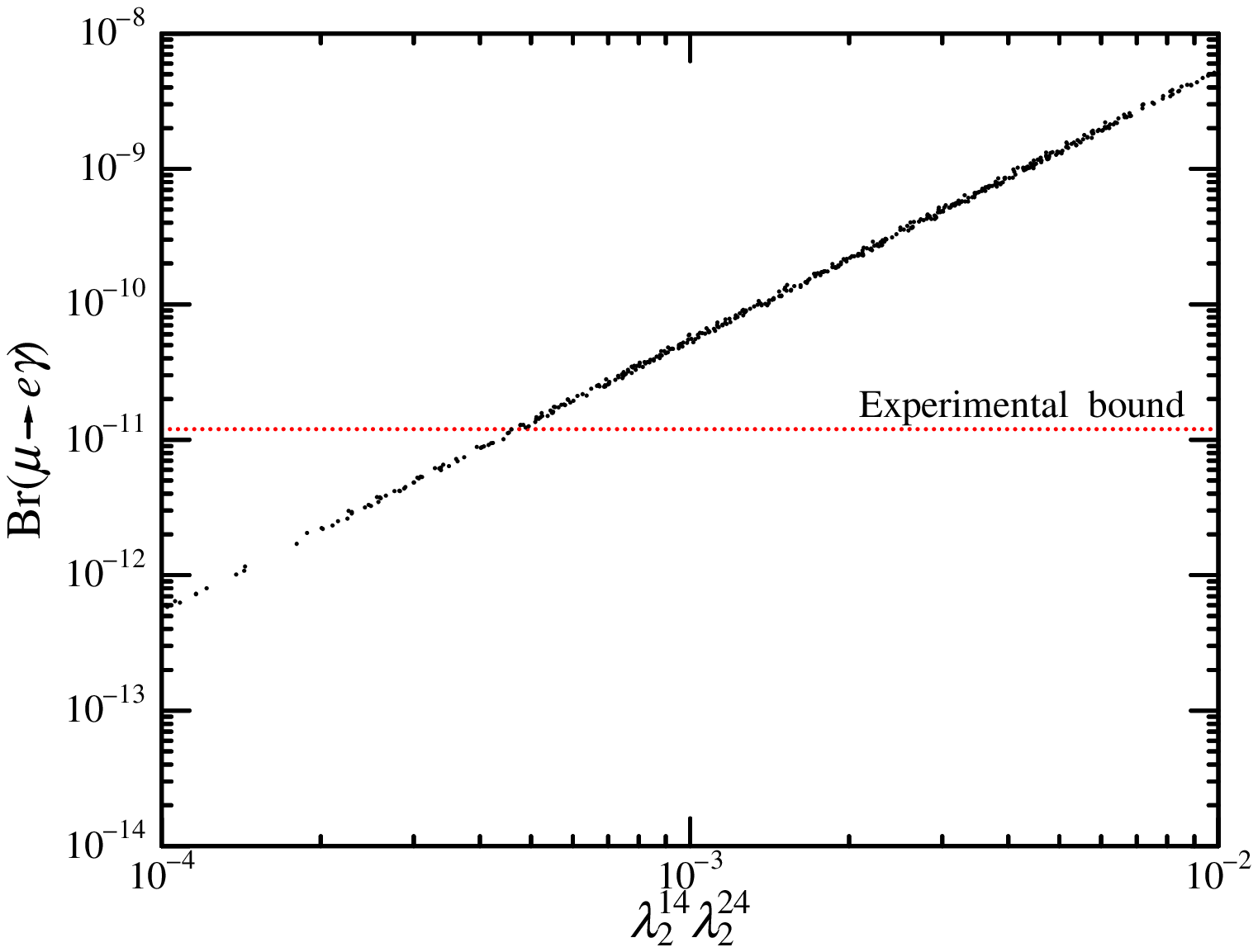}
\end{center}
\caption{
The relation between Br($\mu \rightarrow e \gamma$)
and $ \lambda_2^{14} \lambda_2^{24} $ is plotted in the model (2)
for $ \tan \beta = 5 $.
The points represent the branching ratio
for random values of $ \lambda_2^{i4} $.
\label{l2}}
\end{figure}

In fig.\ref{l2}, Br($\mu \rightarrow e \gamma$) is shown
as a function of $ \lambda_2^{14} \lambda_2^{24} $
by taking randomly the values of $ \lambda_2^{i4} $.
We have also taken typically $ \tan \beta = 5 $,
$ v_3 = 500 {\rm GeV} $, $ m_0 = 300 {\rm GeV} $, $ a_0 = 1 $,
$ m_1 = 300 {\rm GeV} $, $ M_L = 100 {\rm GeV} $,
$ \lambda_3 = 0.8 $ and $ \lambda_4 = 0.7 $.
It is clearly observed in fig.\ref{l2}
that Br($\mu \rightarrow e \gamma$) is roughly proportional
to $ ( \lambda_2^{14} \lambda_2^{24} )^2 $,
as indicated in eq.(\ref{52}).
It is interesting to observe in fig.\ref{l2}
that $ {\rm Br} ( \mu \rightarrow e \gamma ) $ becomes comparable
to the experimental bound $1.2 \times 10^{-11}$
\cite{brbound}
for the rather small ordinary-exotic couplings
$ ( \lambda_2^{14} \lambda_2^{24} )^{1/2} \sim 0.02 $
with the moderate value of $ \tan \beta \sim 5 $.

\begin{figure}
\begin{center}
\includegraphics[height=9cm]{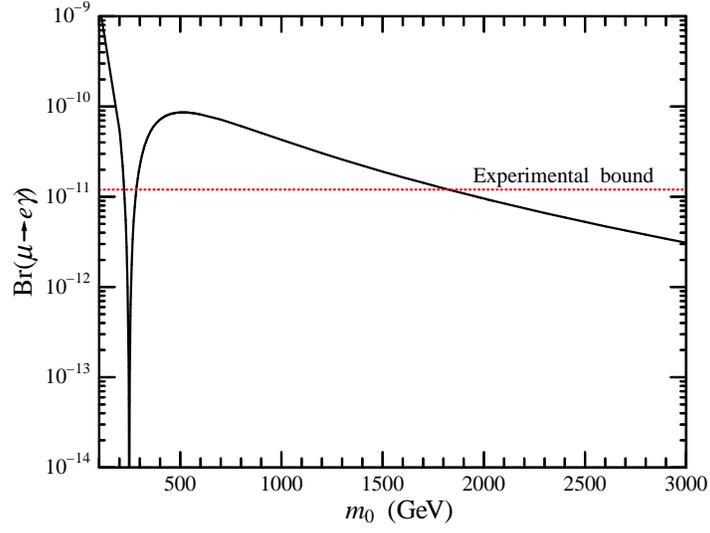}  
\end{center}
\caption{
The relation between Br($\mu \rightarrow e \gamma$)
and the gravitino mass $m_0$ is plotted in the model (2)
for $m_0 \leq 3 \ {\rm TeV}$ with $ \lambda_2^{i4} = 0.07 $
and $ \tan \beta = 5 $.
\label{largem0}}
\end{figure}

The relation between Br($\mu \rightarrow e \gamma$)
and the gravitino mass $m_0$ is shown in fig.\ref{largem0}
for $m_0 \leq 3 {\rm TeV}$ with somewhat larger $ \lambda_2^{i4} = 0.07 $,
where the other parameters are taken typically as
$ \tan \beta = 5 $, $ v_3 = 500 {\rm GeV} $, $ a_0 = 1 $,
$ m_1 = 1000 {\rm GeV} $, $ M_L = 100 {\rm GeV} $,
$ \lambda_3 = 0.8 $ and $ \lambda_4 = 0.7 $.
It is here noticed that cancellation between
the neutralino and chargino contributions occurs for certain values
of $m_0$, where Br($\mu \rightarrow e \gamma$) becomes substantially small.
It is also remarkable in the model (2)
that even for the relatively large soft supersymmetry breaking
with $ m_0 \sim 1 \ {\rm TeV} $,
Br($\mu \rightarrow e \gamma$) can be comparable to the experimental bound
by taking the relatively large ordinary-exotic lepton couplings
$ \lambda_2^{i4} \sim 0.1 $.

We have also the supersymmetric contributions
to $ \tau \rightarrow e \gamma $ and 
$ \tau \rightarrow \mu \gamma $, which are estimated in eq.(\ref{51})
by replacing $ m_\mu $ with $ m_\tau $
and the $ \lambda_2^{i4} $ couplings.
The resultant branching ratios of $ \tau \rightarrow e \gamma $ and
$ \tau \rightarrow \mu \gamma $ are similar to
that of $ \mu \rightarrow e \gamma $,
since they are almost independent of $ m_\tau $ in this case.
Therefore, by considering the experimental bounds on the branching ratios,
the $ \mu \rightarrow e \gamma $ seems to be more promising
to observe the lepton flavor violation in the model (2)
with the bare mass term, if the ordinary-exotic lepton couplings
$ \lambda_2^{i4} $ are comparable each other.

The possible contributions of the $ \lambda_2 $ coupling
to the flavor violation in the quark sector are estimated
in this model (2) as follows.
We have here the significant left-handed squark mixings such as
$ ( M^{2 \prime}_{{\tilde{\cal D}}_{{\rm L}{\rm L}}} )_{12} / m_0^2
\sim \lambda_2^{14} \lambda_2^{24} $,
which is the same as the left-handed slepton mixings
given in eq.(\ref{50}).
Then, a constraint $ \lambda_2^{14} \lambda_2^{24} \lesssim 10^{-2} $
is placed from the $ K^0 $-$ {\bar K}^0 $ system \cite{Gabbiani:1996}.
This constraint is, however, less stringent
than that from the $ \mu \rightarrow e \gamma $.

\section{Summary}
In summary, we have investigated the lepton flavor violating processes,
especially the $\mu \rightarrow e \gamma$ decay,
which are obtained from
the mixing between the ordinary leptons and the vectorlike leptons.
Although the lepton FCNC's appear at the tree level,
their effects are small naturally because of 
the hierarchy of the charged lepton masses.
Hence, the fine tuning of parameters is not necessary
to suppress the FCNC's sufficiently.
In the supersymmetric model, significant effects of lepton flavor violation
are obtained from the slepton mixing which is generated
by radiative corrections with the ordinary-exotic lepton couplings.
In the model (1) without the bare mass term
of the vectorlike lepton doublets,
the supersymmetric contributions to the $\mu \rightarrow e \gamma$ decay
are rather suppressed by $ m_e $ as well as $ m_\mu $
due to the approximate $ {\rm U(1)}_e \times {\rm U(1)}_\mu $ symmetry,
which is similar to the gauge boson mediated contributions.
It is, however, remarkable that these supersymmetric
contributions to $ {\rm Br} ( \mu \rightarrow e \gamma ) $ are
substantially enhanced by the factor $ \tan^6 \beta $
compared to the gauge boson mediated contributions.
Then, if $ \tan \beta $ is larger than 20,
$ {\rm Br} ( \mu \rightarrow e \gamma ) $
might be comparable to the present experimental bound.
In the model (2) with the bare mass term,
much larger contributions to $ {\rm Br} ( \mu \rightarrow e \gamma ) $
are obtained through the slepton mixing.
In either case we have shown that the slepton mixing contributions
to the $\mu \rightarrow e \gamma$ decay can be large
enough to be observed in the near future experiments.
It should also be noted that in the model (1)
the $\tau \rightarrow \mu \gamma$ decay may be more promising
as seen in eq.(\ref{48}).

The discovery of $\mu \rightarrow e \gamma$
clearly indicates the new physics.
The supersymmetric contributions to the $\mu \rightarrow e \gamma$ decay
have been considered so far in the literature
for other supersymmetric models such as
the SUSY GUT \cite{Barbieri:1994pv,Barbieri:1995tw,hisano2}
and the MSSM with right-handed neutrinos
\cite{Borzumati:1986qx,hisano,Hisano:1998fj}.
The predictions of Br($\mu \rightarrow e \gamma$)
in these models are within reach of the future experiments.
These are, however, the effects of ultra high energy physics
much above the electroweak scale.
In the supersymmetric model with vectorlike leptons,
which may be motivated by the ${\rm E_6}$ type unification,
the significant contributions to the $\mu \rightarrow e \gamma$ decay
can be obtained due to the exotic leptons and their scalar partners
which exist just around the electroweak scale.
Therefore, the direct experimental search is feasible
for these new particles \cite{l3,search}.
It is also observed in the model (2) with the bare mass term
of vectorlike leptons that even if the soft supersymmetry breaking scale
is around $ 1 \ {\rm TeV} $, Br($\mu \rightarrow e \gamma$)
can be comparable to the current experimental bound
with relatively large ordinary-exotic lepton couplings.
These features are salient to the supersymmetric model
with vectorlike leptons in contrast to the other supersymmetric models.

\acknowledgements
We would like to thank
Y. Okada and J. Hisano for useful discussions.

\newpage
\begin{figure}
\begin{center}
\includegraphics*[0.3cm,0cm][7.5cm,7cm]{neutralino.ps}
\includegraphics*[0.3cm,-0.3cm][7.5cm,7cm]{chargino.ps}
\caption{One-loop diagrams for
the $\mu \rightarrow e \gamma$ process
mediated by the slepton mixing.\label{meg}}
\end{center}
\end{figure}

\end{document}